\documentclass[12pt,a4paper]{article}
\usepackage{a4wide}
\usepackage{graphicx}

\def\beq{\begin{equation}}
\def\eeq#1{\label{#1}\end{equation}}
\def\eeqn{\end{equation}}

\def\beqa{\begin{eqnarray}}
\def\eeqa#1{\label{#1}\end{eqnarray}}
\def\eeqan{\end{eqnarray}}

\let\bar=\overbar

\def\Dslash{\not{\hbox{\kern-4pt $D$}}}
\def\dslash{\not{\hbox{\kern-2pt $\del$}}}

\def\msb{{\bar{\ssstyle M \kern -1pt S}}}

\def\Title#1{\begin{center} {\Large {\bf #1} } \end{center}}

\begin{document}

\Title{DEAP-3600 dark matter experiment}

\bigskip\bigskip


\begin{raggedright}  
{\it Nasim Fatemighomi  \footnote{Department of Physics,
Royal Holloway University of London,  Egham, UK} for the DEAP-3600 collaboration \index{Reggiano, D.}\\
}
\small
\begin{center} 
Presented at: PIC 2015, Physics in Collision, \\ 15-19 September 2015 \\
University of Warwick

\end{center}

\bigskip\bigskip
\end{raggedright}

\section*{Abstract} 

DEAP-3600 is a single phase liquid argon (LAr) dark matter experiment,
located  2 km underground at SNOLAB, in Sudbury, Canada.  The detector
has 1 tonne fiducial mass of LAr.  The target sensitivity  to
spin-independent scattering of 100 GeV weakly interacting massive
particles (WIMPs) is 10$^{-46}$ cm$^{2}$.   The DEAP-3600 background
target is $<$ 0.6 background events in the WIMP region of interest in
3 tonne-years.  The strategies to achieve this background include pulse
shape discrimination to mitigate electron recoil and using ultra low
radioactive materials for  detector  construction.  Furthermore, to
reduce neutron and alpha backgrounds,  the DEAP-3600 acrylic vessel
was sanded in situ   to mitigate radon exposure of surfaces during construction and fabrication.    The experiment is currently in the commissioning phase and will begin  physics data taking later this year.  This paper presents an overview of the experiment, its cross-section sensitivity to WIMPs and its current status.

\section{Introduction}
The Standard Model of particle physics can only describe 15.5\% of the
matter in the Universe. Numerous cosmological observations indicate
that the remaining 84.5\% does not interact electromagnetically and
therefore is known as dark matter~\cite{plank}. A good candidate for dark matter would be a
particle that interacts gravitationally and is stable or long-lived
enough to have survived since the Big Bang. It must be
non-realistivistic, non-baryonic and electrically neutral. The most
favourable dark matter candidates, which are predicted by models beyond
the Standard Model, are the weakly interacting massive particles
(WIMPs). The search for dark matter has become one of the top
priorities in the particle physics community. The observable in a
labratory-based experiment which measures scattering interaction
between WIMPs and target nuclei, is recoil energy and rate of interactions.  

The DEAP-3600 project is one of several direct detection experiments
worldwide,  using a noble liquid as target and detector
media. DEAP-3600 is a single phase liquid argon (LAr) dark matter detector,
measuring the scintillation signal produced from energy deposition. LAr scintillation has a powerful pulse shape
discrimination (PSD) property allowing efficient identification of
nuclear recoils, from WIMP interaction,  in the presence of electron recoil background.
\section {The DEAP-3600 detector at SNOLAB}
\begin{figure}[h]
\centering
\includegraphics[width=7.0cm]{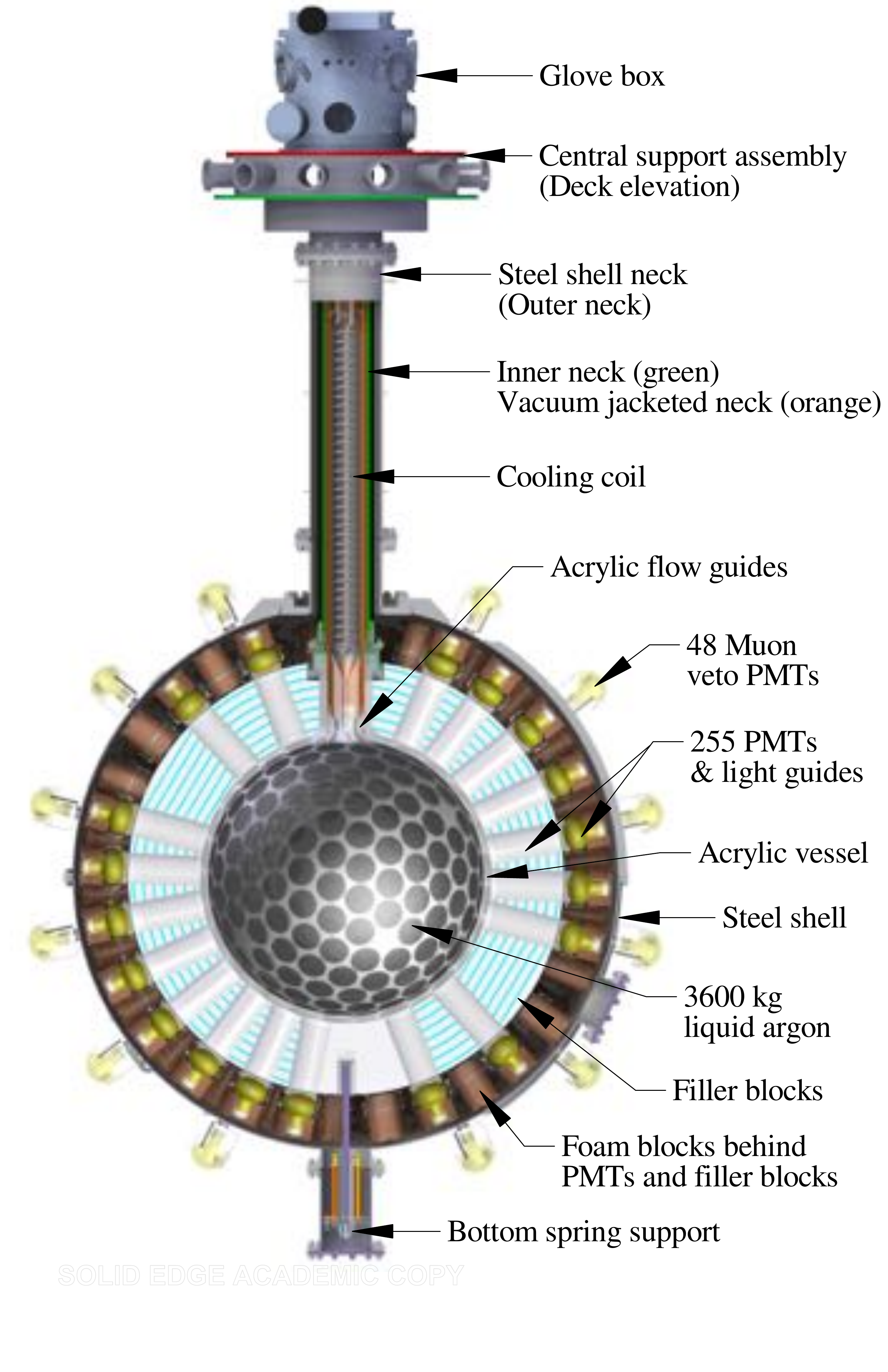}
\caption{A schematic diagram of the DEAP-3600 detector}
\label{fig-1}
\end{figure}

DEAP-3600 is 
situated  2 km underground at SNOLAB, Sudbury, Ontario.  It is
composed of  an 85~cm radius ultra-clean acrylic vessel (AV) which
will be filled with 3600~kg LAr.  The inside of the AV is coated with the
wavelength shifter  tetraphenyl
butadiene (TPB), which shifts the ultraviolet (UV)
light generated by argon scintillation to the visible blue
region. Bonded to the AV are  255 acrylic light guides,  each coupled
to a  high quantum efficiency photomultiplier tube (Hamamatsu R5912).
The empty space between the light guides  is filled with polyethylene
filler blocks. The filler blocks and the light guides provide neutron
shielding and thermal insulation.   A    spherical stainless steel
shell encloses all the internal components and is instrumented with
outward-looking muon veto photomultiplier tubes (PMTs).  The entire assembly is housed in an 8 meter diameter water shield tank.   The detector is cooled through  a liquid
nitrogen  filled  cooling  coil,  installed  in  the  neck  with  acrylic  flow  guides  attached
to  the  bottom.  A schematic view of the DEAP-3600 detector is given in Figure~\ref{fig-1}.

In order to reduce the risk of accidental contamination,  radioactive
calibration sources  were developed to be used outside of the AV.  The
water tank is equipped with three vertical calibration tubes that will
allow periodic deployment of an Am-Be neutron source at the
equator of the steel shell. Gamma calibrations will be performed with
a  $^{22}$Na source moved through an additional circular tube
installed around the steel shell. In addition,  internal and external
optical calibration systems were developed in order to characterize the
detector response. The internal optical calibration system (triggered
laser ball system) was  deployed for a short time to the centre of the
AV before filling the detector with argon.  The purpose of the system
is to measure the PMT relative timing response and efficiency,  as
well as, TPB and light guides optical properties. It  consists of multi-wavelength laser pulser, an optical fibre and a Perfluoroalkoxy polymer (PFA) flask filled with light diffusing material.  The external optical calibration systems consist of 20 LED driven optical fibres for light injection, attached to selected light guides. 

A unique feature of DEAP-3600 is its excellent PSD against electronic recoil events.  The expected electronic background reduction is more than factor of $10^{10}$. The PSD on the LAr scintillation signal is demonstrated with the DEAP-1 prototype experiment~\cite{deap-1}.   The effectiveness of PSD strongly depends on the light yield.  The projected DEAP-3600 light yield is 8~pe/keV.   This  high light yield is due to high PMT coverage (79\%) and running the detector in single phase.  Unlike time projection chambers, the ionization signal is not collected in single phase dark matter detectors. This leads to no scintillation light loss from  ions drifting in  electric field. 
\section{Projected background and sensitivity}
Cosmogenic backgrounds are mitigated by the depth of SNOLAB and water \v{C}erenkov veto. Ultra-low background techniques were used in order to aim for negligible radioactive background events in the WIMP signal region.  The expected number of background events in the WIMP region of interest is 0.6 in 3 tonne-years from all sources~\cite{background}.   

The dominant source of electron recoil background events is from
$^{39}$Ar beta decay (1~Bq/kg).  The  PSD  on the scintillation signal
will be used to mitigate this background.  Low background gamma assay,
radon emanation measurements and alpha counting systems at SNOLAB and
Queen's University  were used  to select low background materials and
eliminate neutron and $\gamma$ backgrounds. The targeted radon emanation from
all the  argon  wetted process system components is less than ~$5~\mu$Bq. 
 Radon in the cryogen  will be removed by a custom-made charcoal radon
 trap installed in the argon process system.    The $^{210}$Pb  level
 in the acrylic, which is a source of background such as ($\alpha$, n)
 neutrons, is measured to be $< 2.2 \times 10^{-19}$
 g/g~\cite{pb-210}.  The AV surface was exposed to radon in air during
 construction. $^{210}$Pb builds on surfaces exposed to radon and
   produces  $^{210}$Po, which is an $\alpha$ emitter. The $\alpha$ particle
 loses its energy in bulk TPB and may mimic WIMP-like events.  To reduce
 surface $\alpha$ background,  the inner surface of the AV was resurfaced
 in situ in order to remove radon daughters.  During and after resurfacing,  the
 AV was  sealed from the lab environment. While purged with low-radon
 gas, the resurfacer   took away 0.4~mm of acrylic. The estimated AV
 event rate after resurfacing is 10~$\alpha$/m$^{2}$/day~\cite{alpha}.
 Neutron, $\gamma$ and $\alpha$ events are further removed by
 fiducialization, while keeping the targeted 1~tonne mass.   
 
\begin{figure}[h]
\centering
\includegraphics[width=8.0cm]{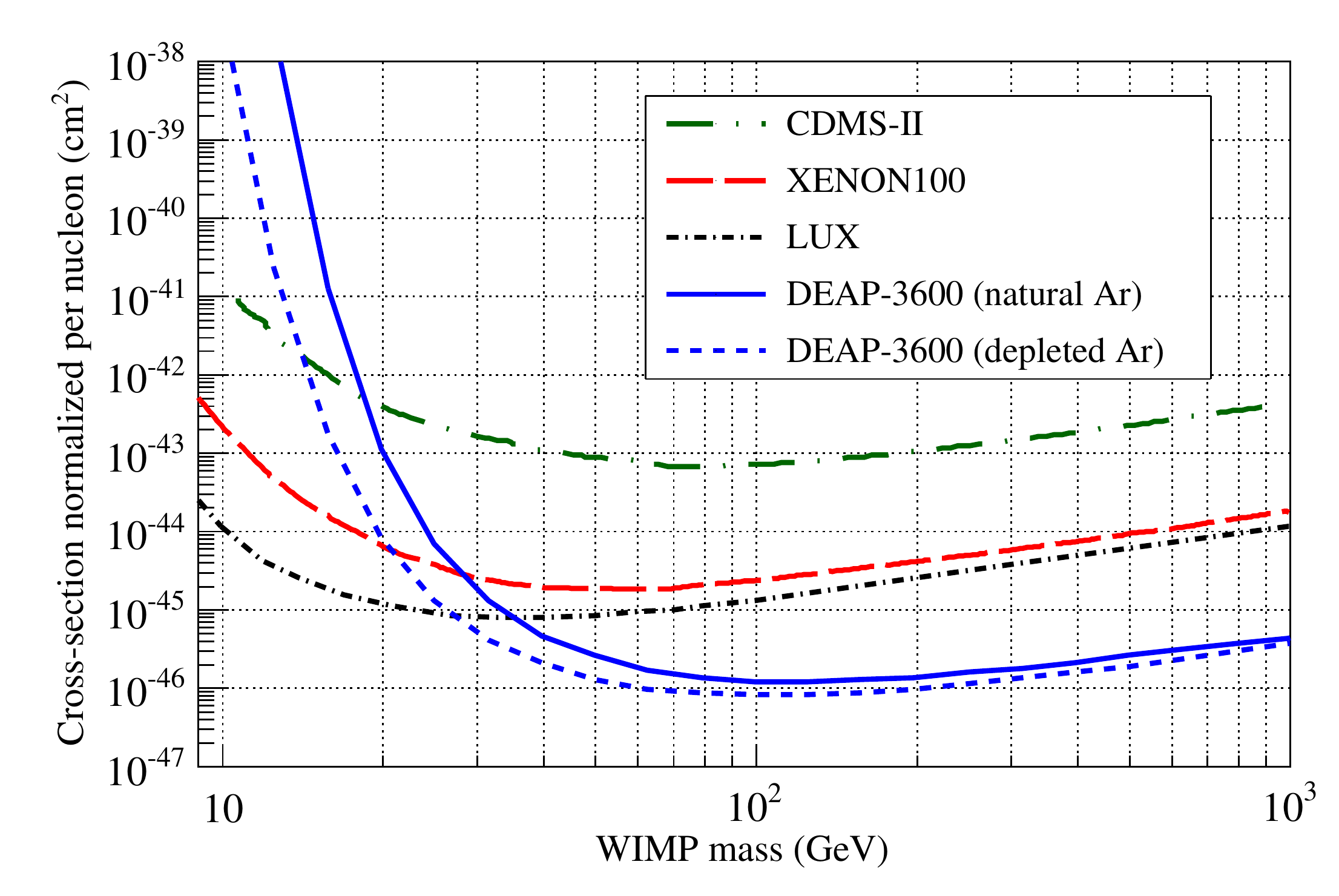}
\caption{DEAP-3600 spin-independent  WIMP-nucleon scattering cross-section sensitivity. For comparison, the current experimental limits from CDMS-II~\cite{cdms}, XENON-100~\cite{xenon} and LUX~\cite{lux} experiments are shown.  }
\label{fig-2}
\end{figure}

DEAP-3600's expected spin-independent WIMP-nucleon cross section sensitivity is $10^{-46}$~cm$^{2}$ for a WIMP mass of 100~GeV. Figure~\ref{fig-2}~\cite{sensitivity} shows the projected DEAP-3600 and current WIMP-nucleon cross section experimental limits. 
\section{Current status and summary}
\begin{figure}[h]
\centering
a)\includegraphics[width=4.05cm]{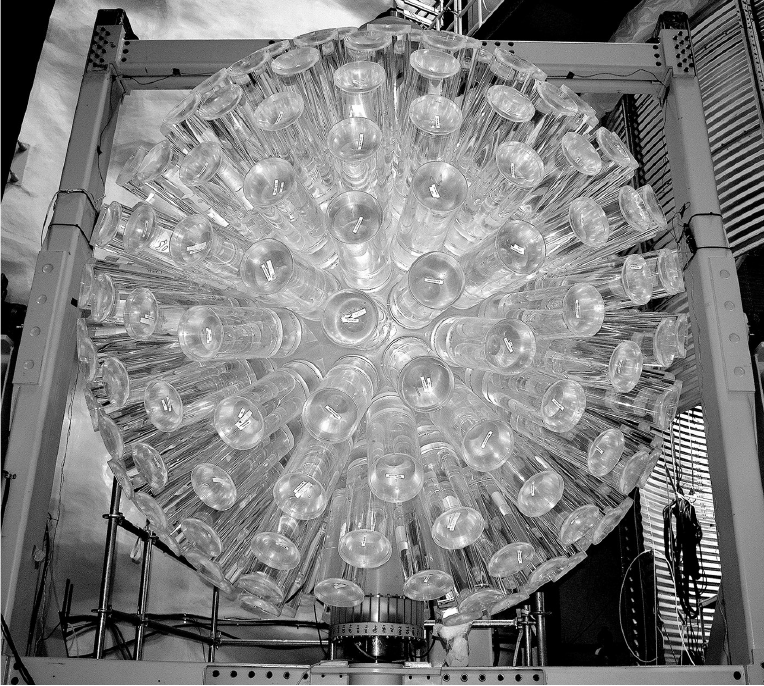}
b)\includegraphics[width=4.45cm]{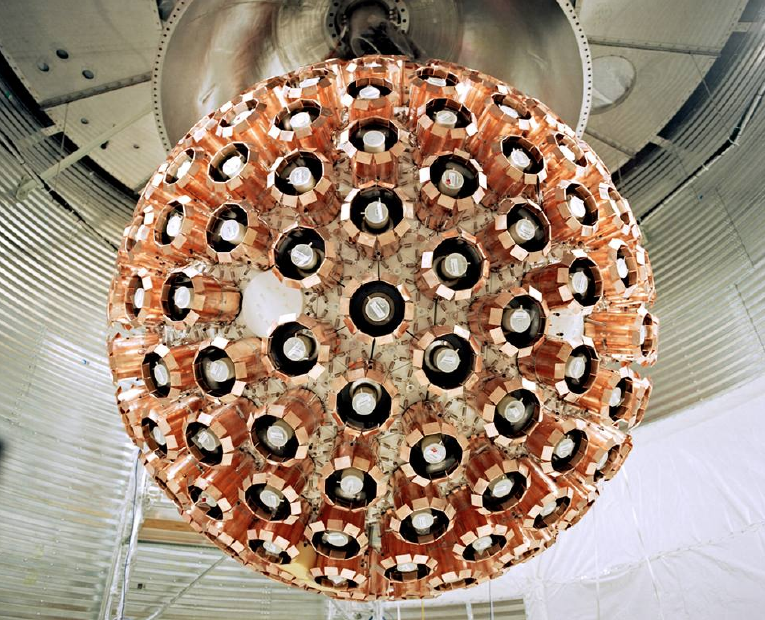}
c)\includegraphics[width=3.3cm]{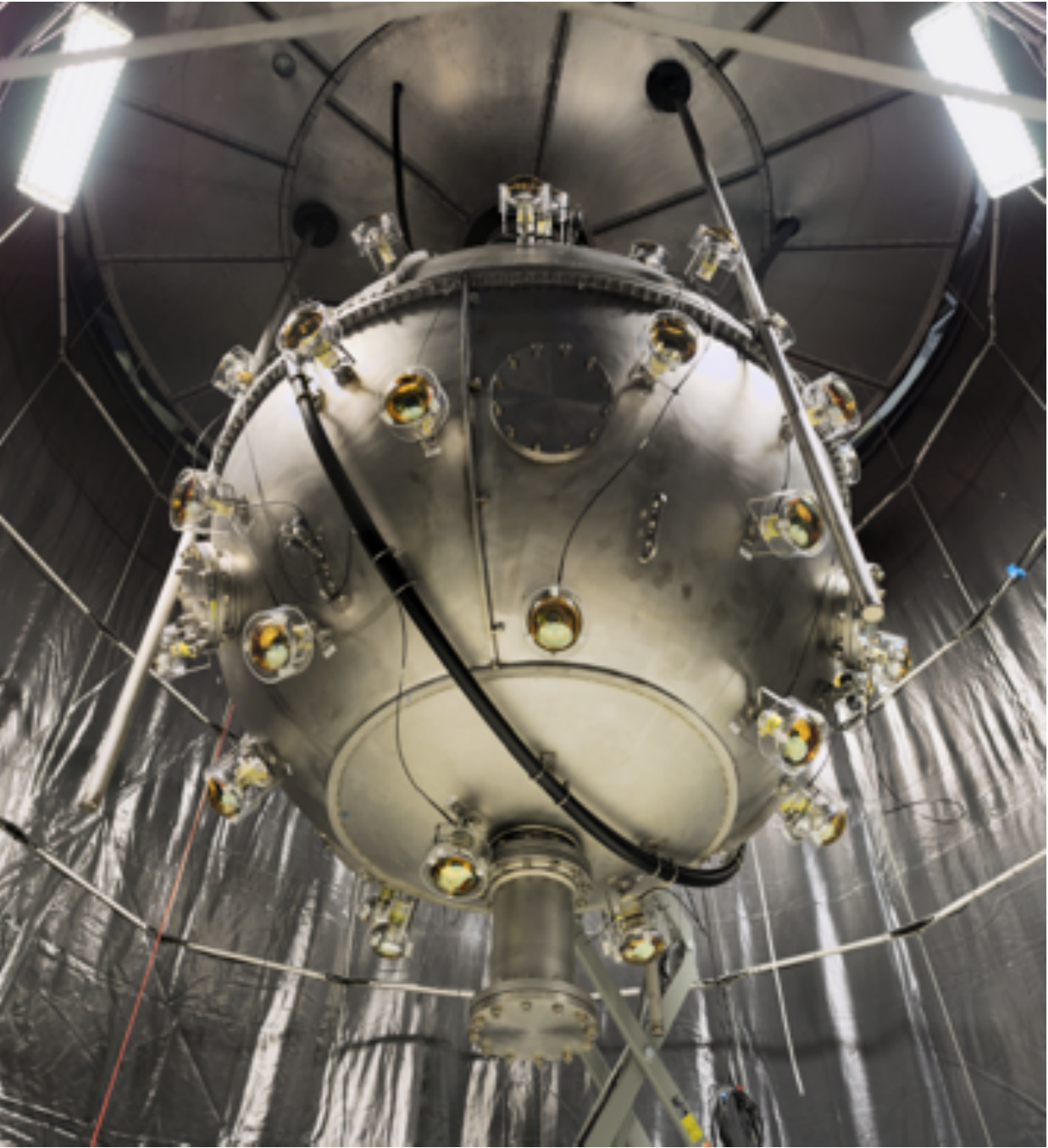}

\caption{a) The DEAP-3600 acrylic vessel attached to an assembly frame after completion of light guides bonding. The full diameter with light guides is 3 m. b)  The DEAP-3600 after  PMTs, reflectors, shielding polyethylene blocks and copper thermal shields installation. c) The DEAP-3600 inside water veto tank after installation of the steel-shell, calibration tubes and veto PMTs. }
\label{fig-av}
\end{figure}

All the major components of the detector are in place.  During 2013,
the AV was fabricated~\cite{avfabrication} and the light guides were bonded to the AV
(Figure~\ref{fig-av}~a). The bonding was performed underground and
included multiple high temperature anneals in low-radon atmosphere. In
2014, the outside of the AV and the  light guides were dressed  with
diffuse and specular reflectors, respectively. The space between the
light guides was filled with polyethylene shielding blocks. The 255
PMTs were installed and coupled to corresponding light guides
(Figure~\ref{fig-av}~b).  In parallel, the process systems were
commissioned.  In early 2015, the steel shell was closed and the calibration tubes and veto PMTs were installed~(Figure~\ref{fig-av}~c).   The electronics, data acquisition system
and external optical calibration  system (LEDs)  have been  commissioned.  

Since February 2015, the detector has been continuously running with the AV under vacuum or filled with ultra-clean N$_{2}$ gas. During this time,  
the data taken during LED runs has been  used to perform PMT charge  calibration. 
 In summer 2015, the laser ball system was deployed in  three
positions inside of the AV. The laser ball was pulsed at 445 nm, 405 nm
and 375 nm wavelengths. The TPB excitation starts at 410~nm
and increases at lower vacuum ultra violet wavelengths~\cite{tpb}. The
reemission photoluminescence spectra peaks at around
425~nm.  Multi-wavelength data allows for an analysis of the full optical properties in the detector. In parallel, the LED 
data (with LED wavelength of 435~nm) can be  used as an independent
cross-check of the optical model and to study the optical stability of
the detector components.  Studies are ongoing to extract the
detector's optical parameters using both laser ball and external LED
data. 


Currently, the water veto tank is sealed and is fully filled. The remaining task is commissioning the detector with Ar gas before
cooling down.  The Ar gas data will be used for Monte  Carlo
simulation tuning, electronic readout adjustments and background studies. 




\section{Acknowledgements}
This work is supported by the National Science and Engineering Research Council of
Canada (NSERC), by the Canada Foundation for Innovation (CFI), by the Ontario
Ministry of Research and Innovation (MRI) and by the European Research Council
(ERC). We thank Compute Canada,  Calcul Qu\'{e}bec, McGill University's centre for
High Performance Computing and the High Performance Computing Virtual Laboratory (HPCVL),  for computational support  and data storage.   We are  grateful  to
SNOLAB and Vale Canada, Ltd.  for excellent on-site support.

\end{document}